\documentclass[sigconf]{acmart}

\usepackage{textcomp}
\usepackage{microtype}
\usepackage{graphicx}
\usepackage{subfigure}
\usepackage{booktabs} 

\usepackage{amsmath}
\usepackage{mathtools}
\usepackage{amsthm}
\usepackage{algorithm}
\usepackage{algorithmic} 
\usepackage[utf8]{inputenc} 
\usepackage[T1]{fontenc}    
\usepackage{url}            
\usepackage{booktabs}       
\usepackage{nicefrac}       
\usepackage{microtype}      
\usepackage{xcolor}         
\usepackage{natbib}
\usepackage{multirow}
\usepackage{colortbl}
\usepackage{amsmath}
\usepackage{graphicx}
\usepackage{tcolorbox}
\usepackage{listings}

\usepackage{scalerel}
\usepackage{tikz}
\usetikzlibrary{svg.path}
  
\usepackage{seqsplit}

\newcommand{\ourapp}{\textsc{IntelliSA}}
\newcommand{\glitch}{\textsc{Glitch}}

\newcommand{\ea}{\textit{et al.}}
\newcommand{\smallsection}[1]{\noindent\underline{\noindent {\bf #1}.}\hspace{1mm}}

\definecolor{mygreen}{rgb}{0.0, 0.5, 0.0}

\newcommand{\revised}[1]{{\color{black}{#1}}}

\newcommand{\rqone}{\textbf{(RQ1) How accurate is our \ourapp~approach in detecting security smells in IaC scripts?}}

\newcommand{\rqtwo}{\textbf{(RQ2) What is the cost-effectiveness of \ourapp~approach for locating security smells in IaC scripts?}}

\newcommand{\rqthree}{\textbf{(RQ3) What is the impact of key design choices in our \ourapp~on the effectiveness of IaC security smell detection?}}

\newcommand*\circled[1]{\tikz[baseline=(char.base)]{
            \node[shape=circle,draw,inner sep=0.5pt] (char) {\small{#1}};}}

\newcommand{\icode}[1]{\texttt{\seqsplit{#1}}}

\AtBeginDocument{%
  }

\setcopyright{acmlicensed}
\copyrightyear{2018}
\acmYear{2018}
\acmDOI{XXXXXXX.XXXXXXX}
\acmConference[Conference acronym 'XX]{Make sure to enter the correct
  conference title from your rights confirmation email}{June 03--05,
  2018}{Woodstock, NY}
\acmISBN{978-1-4503-XXXX-X/2018/06}

\begin{document}

\title{IntelliSA: An Intelligent Static Analyzer for IaC Security Smell Detection Using Symbolic Rules and Neural Inference}


\author{Qiyue Mei}
\orcid{0009-0003-8506-6745}
\email{qiyue.mei@student.unimelb.edu.au}
\affiliation{%
  \institution{The University of Melbourne}
  \city{Melbourne}
  \state{Victoria}
  \country{Australia}
}

\author{Michael Fu}
\orcid{0000-0001-7211-3491}
\email{michael.fu@unimelb.edu.au}
\affiliation{%
  \institution{The University of Melbourne}
  \city{Melbourne}
  \state{Victoria}
  \country{Australia}
}

\renewcommand{\shortauthors}{Mei et al.}

\begin{abstract}
    Infrastructure as Code (IaC) enables automated provisioning of large-scale cloud and on-premise environments, reducing the need for repetitive manual setup. However, this automation is a double-edged sword: a single misconfiguration in IaC scripts can propagate widely, leading to severe system downtime and security risks. Prior studies have shown that IaC scripts often contain security smells—bad coding patterns that may introduce vulnerabilities—and have proposed static analyzers based on symbolic rules to detect them. Yet, our preliminary analysis reveals that rule-based detection alone tends to over-approximate, producing excessive false positives and increasing the burden of manual inspection. In this paper, we present \ourapp, an intelligent static analyzer for IaC security smell detection that integrates symbolic rules with neural inference. \ourapp~applies symbolic rules to over-approximate potential smells for broad coverage, then employs neural inference to filter false positives. While an LLM can effectively perform this filtering, reliance on LLM APIs introduces high cost and latency, raises data governance concerns, and limits reproducibility and offline deployment. To address the challenges, we adopt a knowledge distillation approach: an LLM teacher generates pseudo-labels to train a compact student model—over 500× smaller—that learns from the teacher’s knowledge and efficiently classifies false positives. We evaluate \ourapp~against two static analyzers and three LLM baselines (Claude-4, Grok-4, and GPT-5) using a human-labeled dataset including 241 security smells across 11,814 lines of real-world IaC code. Experimental results show that \ourapp~achieves the highest F1 score (83\%), outperforming baselines by 7–42\%. Moreover, \ourapp~demonstrates the best cost-effectiveness, detecting 60\% of security smells while inspecting less than 2\% of the codebase.
\end{abstract}

\ccsdesc[500]{Software and its engineering~Software configuration management and version control systems}  
\ccsdesc[300]{Security and privacy~Software security engineering}

\keywords{Infrastructure as Code (IaC), Security Smells, Static Analysis}



\maketitle

\section{Introduction}
\label{sec:introduction}
Infrastructure as Code (IaC) automates the provisioning and deployment of cloud‑based and containerized environments, where managing thousands of instances manually would be prohibitively time‑consuming and error‑prone. Instead of relying on repetitive manual steps, IaC enables teams to define infrastructure in scripts that can be executed consistently and at scale.
Leading platforms such as Puppet, Ansible, and Chef have become central to this practice, each offering frameworks that simplify configuration and accelerate delivery pipelines.
For instance, Puppet enabled Daiwa Capital to reduce provisioning from weeks to just 10 minutes \cite{puppet_case_study}; Ansible allowed NASA to cut deployment times from hours to minutes \cite{ansible_case_study}; and Chef helped iManage accelerate application delivery by 66\%, reducing customer onboarding from 45 to 15 days \cite{chef_case_study}.

While IaC delivers speed, consistency, and scalability, it also introduces new risks: because infrastructure is defined and executed as code, even small defects or misconfigurations can propagate rapidly across entire systems. Unlike traditional manual errors, which may be isolated, IaC defects are automated and repeatable, meaning a single mistake can instantly affect thousands of resources or critical services. For example, in 2024, a misconfiguration in UniSuper’s IaC provisioning scripts on Google Cloud triggered the deletion of their entire subscription, causing more than a week of downtime and disrupting services for over 620,000 members \cite{unisuper_case_study}.

To address security challenges in Infrastructure as Code (IaC), Rahman~\ea~\cite{rahman2019seven,rahman2020gang,rahman2021different} developed a taxonomy of insecure coding practices, known as security smells, identified through empirical studies and manual analysis of IaC scripts. Building on this taxonomy, they introduced static analyzers such as \textsc{SLIC}, which uses symbolic rule matching to detect security smells in Puppet, and \textsc{SLAC}, which extends this approach to Ansible and Chef. More recently, Saavedra \ea~\cite{saavedra2022glitch} proposed \glitch, a polyglot static analyzer that overcomes the platform-specific limitations of SLIC and SLAC by employing an intermediate representation and symbolic rule matching, achieving state-of-the-art performance in cross-platform smell detection.
However, our preliminary analysis indicates that, similar to other static analyzers for source code, \glitch~suffers from high false‑positive rates due to over‑approximation of rule matches.
\textbf{This limitation highlights the fundamental challenge of fully relying on symbolic rule matching to detect security smells in IaC, as it can lead to inaccurate detection and overwhelm security engineers with excessive false alarms.}

To address the limitation, we propose \ourapp, an \underline{Intelli}gent \underline{S}tatic \underline{A}nalyzer that integrates symbolic rule matching with neural inference. Our framework first performs symbolic static analysis to achieve high recall and broad coverage of security smells.
To mitigate false positives inherent to rule-based detection, we leverage a large language model (LLM) as a teacher model to assess code context and filter out false positive matches.
Although an LLM is capable of performing this filtering, depending on external LLM APIs incurs substantial financial and performance overheads, poses challenges for data governance, and restricts reproducibility and offline usability.
To address this, we leverage the LLM’s false positive detections as pseudo-labels to train a compact classification student model, without the need for ground-truth supervision.
This teacher–student distillation yields a lightweight neural inference component.
Finally, \ourapp~combines symbolic rule matching for broad coverage with distilled neural inference to effectively filter false detections.

We conduct experiments to evaluate \ourapp~against three static analyzers (\textsc{SLIC}, \textsc{SLAC}, and \glitch) and three LLMs (Claude‑4, Grok‑4, and GPT‑5), comparing their effectiveness in detecting security smells in IaC scripts. Our evaluation spans three widely used technologies—Puppet, Ansible, and Chef—and is based on a human-labelled dataset of 241 IaC scripts comprising 11,814 lines of code and nine categories of security smells. Through this extensive study, we address the following three research questions:
\begin{itemize}
\item {\rqone} \\
\textbf{Results.}
\ourapp~outperforms static analysis and LLM baselines, achieving the best F1 scores of 85\%, 88\%, and 77\% on Puppet, Ansible, and Chef, respectively.

\item {\rqtwo}\\
\textbf{Results.}
On Puppet, Ansible, and Chef, our \ourapp~outperforms other baselines on cost‑effectiveness measures, achieving Effort@60\%Recall of 1.14\%, 0.74\%, and 1.66\%, and F1@1\%LOC of 65\%, 85\%, and 55\%, respectively.

\item {\rqthree}\\
\textbf{Results.}
For detecting false positives from a static analyzer, Claude‑4 provided the best balance among LLM teachers, achieving the highest F1 of 89\%. Additionally, CodeT5p proved to be the most effective student model with a macro‑F1 of 79\%, and the 220M parameters matched the 770M variant model’s performance while being 71\% smaller, confirming our efficiency‑oriented design choice.
\end{itemize}

\indent\smallsection{Novelty \& Contributions}
To the best of our knowledge, the primary contributions of this paper are as follows:
\begin{enumerate}
    \item[(1)] We present \ourapp, an intelligent static analyzer that integrates symbolic rule matching with neural inference for detecting security smells, effectively addressing the high false-positive rates of traditional static analyzers.  
    \item[(2)] Through extensive experiments, we demonstrate that our approach achieves higher accuracy and cost-effectiveness compared to existing static analyzers and zero-shot LLM-based methods.  
    \item[(3)] We validate the design choices and provide empirical support for the rationale behind our approach.
\end{enumerate}

\indent\smallsection{Open Science}
\revised{To support open science, we release all resources, including the dataset, processing scripts, experimental scripts, model checkpoints, and the deployed \ourapp~CLI tool—at \url{https://github.com/ColeMei/intellisa}.}

\section{Background \& Related Work}
\label{sec:background}
In this section, we present the background of IaC and the notion of security smells in IaC. We then introduce the nine security smells examined in this paper and present prior work that applies static analysis to detect such smells. Finally, we describe the state‑of‑the‑art static analyzer \glitch, followed by a motivating example and preliminary analysis that highlight its limitations.

\subsection{Infrastructure as Code}
We introduce three widely used Infrastructure as Code (IaC) technologies—Puppet, Ansible, and Chef—which have been shown to significantly reduce deployment times in prior case studies \cite{puppet_case_study,ansible_case_study,chef_case_study}.
Specifically, Puppet manifests, authored in a domain-specific language (DSL) using .pp files, declare resources with syntax like \texttt{``type { 'title': attribute => value }''} to specify desired states, while variables are assigned via \texttt{``=''} and organized into reusable classes for modularity. Ansible playbooks, structured in human-readable YAML (.yml files), outline plays that target hosts and sequence tasks through modules with key-value pairs (e.g., \texttt{``module: key: value''}), enabling straightforward orchestration with variables for parameterization. Chef recipes, embedded in Ruby code (.rb files) and bundled in cookbooks, define resources via blocks such as \texttt{``type 'name' do attribute value end''} to enforce idempotent changes, incorporating Ruby constructs like conditionals and loops for procedural flexibility.



\subsection{Security Smells in IaC Scripts}
A code smell refers to a recurrent coding pattern that signals potential maintenance issues, though it may not always lead to adverse effects and thus warrants scrutiny as an early indicator of deeper problems \cite{fowler2018refactoring}.
Building on the concept of code smells, security smells are recurring patterns in code that hint at underlying security weaknesses requiring further examination; unlike vulnerabilities—which are exploitable flaws causing tangible harm \cite{rahman2021different}—security smells identify risks without specifying exploitation mechanisms.
To contextualize their potential impact, security smells can be linked to entries in the Common Weakness Enumeration (CWE), a standardized taxonomy of software weakness types that provides a bridge between recurring patterns in IaC and recognized classes of security flaws.
Rahman \ea~\cite{rahman2019systematic} first identified gaps in research on IaC defects and security, leading to the proposal of seven security smells in Puppet \cite{rahman2019seven}, later replicated and expanded with two additional smells in Ansible and Chef \cite{rahman2021security}.

Following the classifications of Rahman \ea~\cite{rahman2021security} and Saavedra \ea~\cite{saavedra2022glitch}, we focus on the nine security smells as the basis for this paper:
(1) \textbf{Admin by default (CWE-250):}
Recurring pattern of granting administrative privileges to default users, violating the principle of least privilege.
\emph{Example:} \icode{Puppet: user \{ 'app': groups => ['sudo'] \}};
(2) \textbf{Empty password (CWE-258):}
Using a zero-length string as a password indicates a weak/absent credential.
\emph{Example:} \icode{Ansible: vars: db\_password: ""};
(3) \textbf{Hard-coded secret (CWE-259 and CWE-798):}
Embedding sensitive values (passwords, usernames, private keys) directly in code.
\emph{Example:} \icode{Chef: default['db']['password'] = "P@ssw0rd!"};
(4) \textbf{Missing default in case statement (CWE-478):}
Case/conditional logic omits a catch-all branch, leaving inputs unhandled.
\emph{Example:} \icode{Puppet: case \$osfamily \{ 'Debian': \{...\} \}};
(5) \textbf{No integrity check (CWE-353):}
Downloading remote content without verifying the checksum or signature.
\emph{Example:} \icode{Ansible: get\_url: url=http://ex.com/pkg.rpm dest=/tmp/pkg.rpm};
(6) \textbf{Suspicious comment (CWE-546):}
Comments that signal defects or insecure shortcuts (e.g., TODO, FIXME, HACK).
\emph{Example:} \icode{\# TODO: temporary insecure rule};
(7) \textbf{Invalid IP address binding (a.k.a.\ Unrestricted IP address) (CWE-284):}
Binding services to \texttt{0.0.0.0}, exposing them to all networks.
\emph{Example:} \icode{Ansible: listen\_address: 0.0.0.0};
(8) \textbf{Use of HTTP without TLS (CWE-319):}
Using plain \texttt{http://} for transfers, enabling interception and tampering.
\emph{Example:} \icode{Ansible: get\_url: url=http://ex.com/file.tgz};
and (9) \textbf{Use of weak cryptography algorithms (CWE-326 and CWE-327):}
Relying on weak hashes like MD5 or SHA-1 for security purposes.
\emph{Example:} \icode{Puppet: file \{ '/tmp/x': checksum => 'md5' \}}.

\vspace{-2mm}
\subsection{Static Analysis for IaC Security Smells}
Early work by Rahman \ea~\cite{rahman2019seven} introduced \textsc{SLIC}, one of the first static analysis tools for detecting security smells in Infrastructure-as-Code (IaC) scripts. \textsc{SLIC} targeted Puppet and relied on a parser to tokenize IaC code and a rule engine to match symbolic patterns corresponding to seven predefined security smells. While effective in demonstrating the feasibility of automated smell detection, \textsc{SLIC} was limited in scope to Puppet and relied heavily on handcrafted rules.
To extend coverage beyond Puppet, Rahman \ea~\cite{rahman2021security} later proposed \textsc{SLAC}, a static analysis tool designed for Ansible and Chef scripts. \textsc{SLAC} expanded the catalog of detectable smells (six for Ansible and eight for Chef), adapting its parsing strategy to the YAML-based structure of Ansible and the Ruby-based syntax of Chef. This broadened applicability across IaC technologies, but both \textsc{SLIC} and \textsc{SLAC} remained fundamentally rule-based, inheriting limitations such as susceptibility to false positives and inconsistencies across IaC technologies, since each tool implemented its rules separately for Puppet, Ansible, and Chef.

To address the limitations of \textsc{SLIC} and \textsc{SLAC}, Saavedra \ea~\cite{saavedra2022glitch} proposed \glitch, a polyglot static analysis framework for IaC security smell detection. Unlike earlier tools, \glitch~introduced a technology-agnostic intermediate representation, enabling consistent detection across Puppet, Ansible, and Chef without duplicating implementations for each platform. This unifying design reduced inconsistencies and made it easier to add new security smells by defining rules once for all IaC technologies.
In a large-scale empirical study, \glitch~demonstrated higher precision and recall than both \textsc{SLIC} and \textsc{SLAC}, establishing it as the state of the art in static analysis for IaC security smells.

\revised{Beyond static analysis, recent studies have applied LLMs to IaC smell detection. War~\ea~\cite{war2024detection} relied on fine-tuning with substantial labeled data, while Vo~\ea~\cite{vo2025harnessing} explored zero-shot prompting but reported limited accuracy.}

\subsection{\glitch: A State-of-the-Art Static Analyzer for IaC Security Smells}
\glitch~is a technology‑agnostic static analysis framework for detecting IaC security smells across Ansible, Chef, and Puppet. It introduces a unified intermediate representation (IR) that normalizes heterogeneous syntaxes into a common hierarchical model. The IR captures projects, modules, unit blocks, and atomic units, preserving the semantic elements needed for analysis, and serves as the basis on which symbolic rules are applied to identify security smells across technologies.

\textbf{Step 1: Intermediate Representation.}
To achieve technology-agnostic analysis, \glitch~first transforms IaC scripts into a unified intermediate representation (IR).
The IR follows a hierarchical structure with four levels: projects, modules, unit blocks, and atomic units.
Projects correspond to folders containing modules and scripts; modules group related functionality (e.g., an Ansible role, a Chef cookbook, or a Puppet module); unit blocks represent individual scripts or grouped resources (e.g., a Puppet class or a Chef recipe); and atomic units capture the smallest actionable elements, such as resource definitions.
Each node may contain \texttt{attributes}, \texttt{variables}, or \texttt{conditions}, with flags (e.g., \texttt{has\_variable}) indicating references between values.
For instance, a simple value in the IR can be represented as \texttt{<value> ::= "string" | 42 | True | <id>}. This abstraction allows heterogeneous syntaxes from Ansible, Chef, and Puppet to be expressed uniformly. 

\textbf{Step 2: Symbolic Rule Traversal.}
Once IaC scripts are transformed into the intermediate representation, Glitch applies a set of symbolic rules to detect security smells. The framework traverses the IR using depth‑first search, visiting each project, module, unit block, and atomic unit in turn. At every node, type‑checking functions (e.g., \texttt{isAttribute}, \texttt{isVariable}, \texttt{isAtomicUnit}) and helper predicates (e.g., \texttt{hasDownload}, \texttt{hasChecksum}, \texttt{isDefault}) are evaluated to determine whether a smell is present. Each node is analyzed independently, and multiple smells may be reported for the same construct. Detection relies on configurable string‑pattern functions, which allow the same rules to be adapted across Ansible, Chef, and Puppet without re‑implementation. This rule‑based traversal constitutes the core of Glitch’s smell detection process.

\begin{figure*}[htbp!]
\includegraphics[width=\linewidth]{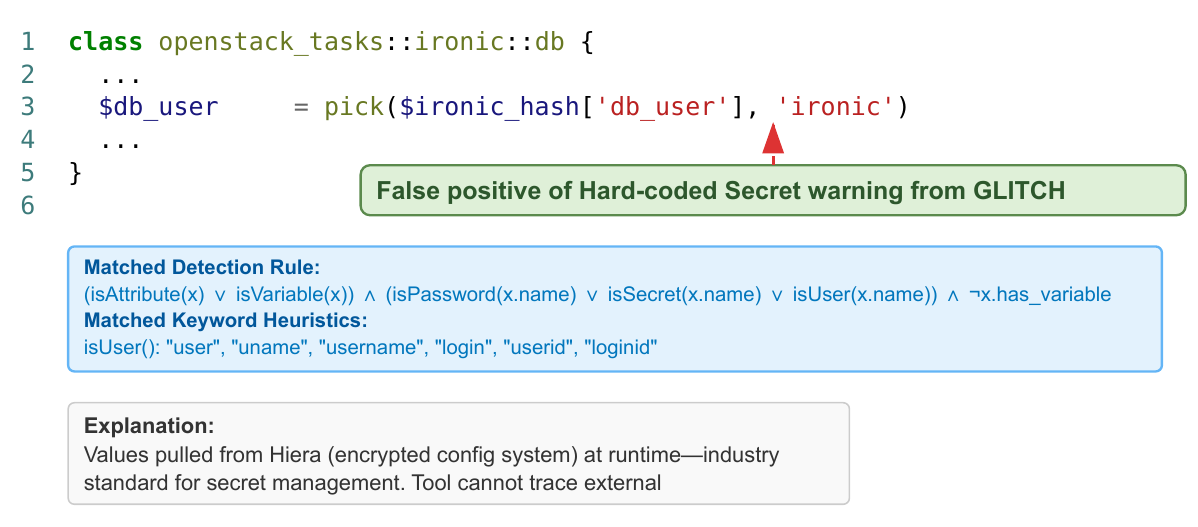}
\caption{(Motivating Example) A false positive case from \glitch: the variable \texttt{\$db\_user} is flagged as a hard-coded secret because its name matches the \texttt{isUser()} predicate and the fallback value \texttt{'ironic'} is a literal. This illustrates how symbolic rule matching without semantic context can misclassify benign defaults as security smells.}
\label{fig:glitch_fp}
\end{figure*}

\smallsection{Motivating Example}
As detailed in Figure~\ref{fig:glitch_fp}, \glitch~first transforms the construct into its intermediate representation (IR), where \texttt{\$db\_user} is identified as a variable node with a fallback literal value. During symbolic rule traversal, the rule matches because the variable name \texttt{db\_user} contains the substring ``user,'' which satisfies the \texttt{isUser()} predicate in \glitch's predefined heuristics, and the fallback value \texttt{'ironic'} is a hard-coded string rather than a variable reference. Consequently, \glitch~flags the assignment as a hard-coded secret. Although the value is merely a benign default database username, the analyzer conservatively reports it as a potential secret. \textbf{This false positive highlights the limitation of fully relying on symbolic rule matching.}

\subsection{Preliminary Analysis \& Limitations}
We conduct a preliminary analysis of \glitch~to empirically assess its effectiveness in detecting security smells in IaC scripts. This analysis is motivated by our observation that \glitch, which relies exclusively on symbolic pattern-matching rules, may generate a substantial number of false positives.
To quantify this limitation, we leverage the oracle dataset provided by Saavedra \ea~\cite{saavedra2022glitch}, which consists of 80 Puppet IaC files labeled by three independent human experts. Consistent with their study, we focus on nine representative security smells described in the preceding section.
We evaluate \glitch~using two standard metrics: precision, which reflects the extent of false alarms, and recall, which captures the tool’s ability to correctly identify security smells.

In addition to static analysis, recent advances in large language models (LLMs) have demonstrated strong capabilities in software development and coding tasks.
Within the security domain, LLMs have been applied to vulnerability detection \cite{steenhoek2024comprehensive,lu2024grace} and even automated vulnerability repair \cite{zhou2024out,yu2025evaluation}.
However, it remains unclear whether such models can be used off the shelf to detect security smells in IaC scripts, despite their general code intelligence.
To explore this question, we design a prompt that specifies the role of a static analyzer, incorporates the formal \glitch~detection rules and keyword‑based heuristics, provides clear task instructions, embeds the raw IaC code input, and defines a structured return template for reporting detected smells.
This ensures the LLM is explicitly guided on the detection task, the applicable rules, and the expected output format.
We then evaluate Claude‑4 using the same oracle dataset and metrics as in our preliminary analysis of \glitch.
Table~\ref{tab:prelim-analysis-results} presents the preliminary analysis results of using \glitch~and Claude-4 for security smell detection.

\subsubsection{\glitch~for IaC Security Smell Detection}
Our preliminary evaluation shows that \glitch~achieves a high recall of 94\%, indicating that symbolic rule matching is effective in capturing many true positive cases. However, precision is substantially lower at 42\%, reflecting a large number of false positives.

\smallsection{Limitation: Over-Approximation in Rule Matching}
The preliminary results suggest that while symbolic rule matching enables broad coverage, it does so at the cost of precision. The over-approximation inherent in \glitch~often flags patterns that syntactically resemble security smells but do not actually embody security weaknesses. This behavior is consistent with our motivating example in Figure \ref{fig:glitch_fp}, where \glitch~reported a false positive despite the absence of a genuine smell.
Consequently, \glitch~produces numerous false positives, limiting its practical utility. This limitation underscores the need for mechanisms that can refine or complement static rule-based detection to reduce false positives and improve the reliability of IaC security smell analysis.

\subsubsection{LLM-Based IaC Security Smell Detection}
Our preliminary evaluation shows that Claude achieves a recall of 62\%, substantially lower than the 94\% achieved by \glitch. Precision is slightly higher at 50\%, yet the overall F1 score of 55\% remains below that of \glitch, indicating weaker overall effectiveness.

\smallsection{Limitation: Under-Approximation in LLM Detection}
These preliminary results indicate that although LLM-based detection reduces some false positives relative to purely symbolic rule matching, this improvement comes at the expense of recall, with a substantial proportion of true security smells left undetected.
The resulting F1 score demonstrates that the trade-off between recall and precision is unfavorable in this setting.
This limitation highlights the difficulty of applying LLMs off the shelf to domain-specific security smell detection without further adaptation or integration with complementary techniques.

\begin{table}[htbp!]
  \centering
  \caption{(Preliminary Analysis) Security smell detection performance on the oracle test set for \textbf{Puppet}. Better performance highlighted in bold.}
  \label{tab:prelim-analysis-results}
  \begin{tabular}{l|ccc}
    \hline
    Method & Precision & Recall & F1 \\
    \hline
    GLITCH & 0.418 & \textbf{0.938} & \textbf{0.578} \\
    \hline
    Claude-4  & \textbf{0.500} & 0.615 & 0.552 \\
    \hline
  \end{tabular}
\end{table}
\begin{figure*}[t]
\includegraphics[width=\linewidth]{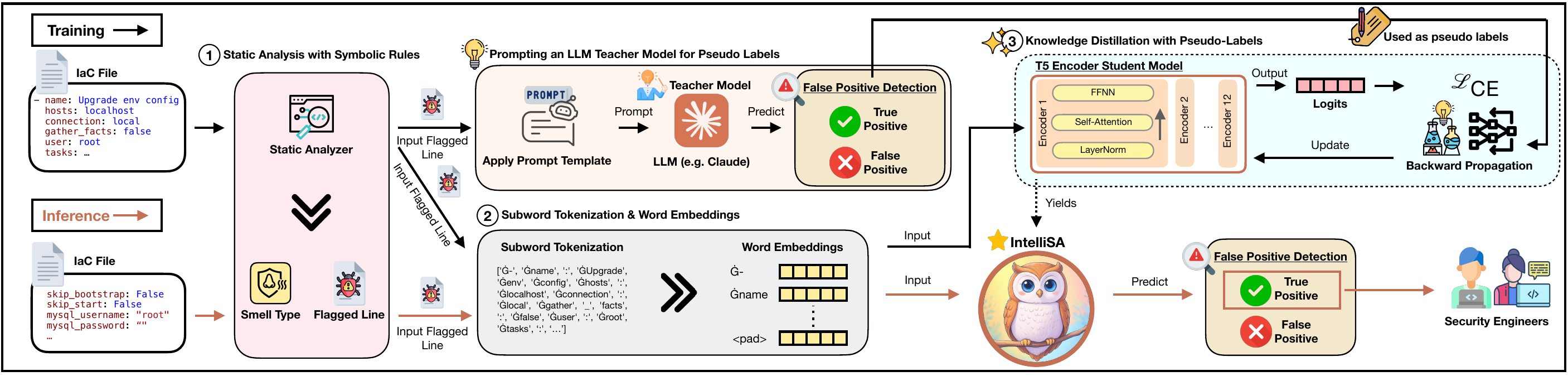}
\caption{Overview of \ourapp~training process and inference usage.}
\label{fig:overview}
\end{figure*}

\section{\ourapp: Framework Overview}
\label{sec:approach}
\textbf{Design Rationale.}
Our preliminary analysis revealed complementary limitations in symbolic and neural approaches to IaC security smell detection: while \glitch~achieves very high recall through symbolic rule matching, its overapproximation produces numerous false positives, whereas LLM-based detection reduces false positives but suffers from underapproximation and diminished recall.
To address the dominant limitation of \glitch, we investigated whether an LLM could serve as a false positive detector on top of \glitch’s warnings. Interestingly, this task proved more tractable than direct smell detection, with an LLM such as Claude achieving over 88\% F1 in identifying false positives—substantially reducing spurious reports while preserving most true positives.
However, reliance on an LLM API introduces high cost and latency, raises data governance concerns, and limits reproducibility and offline deployment—making them impractical for security-sensitive environments where local control and data integrity are essential.
Training smaller domain-specific models with human-labeled data could mitigate this challenge, but manual labeling is prohibitively expensive: for instance, Saavedra \ea~\cite{saavedra2022glitch} required seven raters to label only 241 oracle scripts out of a corpus of 200k, underscoring the impracticality of large-scale annotation.
To overcome these barriers, we propose \ourapp, a framework that leverages LLMs to generate pseudo‑labels for false positive detection without requiring human annotation. We treat a large LLM (e.g., Claude, estimated at over 100B parameters) as a teacher model and distill its labeling behavior into a compact 200M‑parameter student model that is more than 500× smaller.
We then integrate the distilled student model with a static analyzer, forming an intelligent system that combines symbolic rule matching with neural inference for IaC security smell detection.
In what follows, we introduce each step in our \ourapp~framework as outlined in Figure~\ref{fig:overview}.

\subsection{Generating Pseudo-Labels with an LLM Teacher}
Let \(x_\text{train}\) denote an IaC file (e.g., an Ansible \texttt{.yml}), and let \(\mathcal{A}\) be a static analyzer (\glitch) that applies symbolic rules to detect candidate security smells.
In Step \circled{1}, we run \(\mathcal{A}\) on \(x_\text{train}\), which yields a (possibly empty) set of warnings \(\mathcal{W}(x_\text{train})=\{w_i\}\), where each \(w_i=(\ell_i, s_i)\) encodes the flagged line \(\ell_i\) and smell type \(s_i\).
For each \(w_i\in\mathcal{W}(x_\text{train})\), we construct a prompt using the chat template that includes the IaC content \(x_\text{train}\), the detected location \(\ell_i\), and the rule context \(s_i\), and query an LLM teacher \(\mathcal{T}\) to infer a binary label \(y_i \in \{\text{TP}, \text{FP}\}\) indicating whether the warning is a true or false positive.
The resulting set of teacher labels \(\mathcal{Y}(x_\text{train})=\{y_i\}\) serves as pseudo-labels for the knowledge distillation stage described in the following section.

\subsection{Knowledge Distillation with Pseudo-Labels}
\label{subsec:ce_loss}
In Step \circled{2}, given a warning \(w=(\ell_i, s_i)\) from the \glitch~output \(\mathcal{W}(x_\text{train})\), we extract the corresponding IaC file \(x_\text{train}\) and focus on the flagged line \(\ell_i\) with its associated smell type \(s_i\). The flagged line alone is treated as the input sequence and tokenized into subword units \(\{t_1, t_2, \ldots, t_m\}\) using a subword tokenizer~\cite{sennrich2015neural}, where \(m\) denotes the sequence length. 

\revised{Each token \(t_j\) is mapped to a 768-dimensional embedding vector \(\mathbf{e}_j \in \mathbb{R}^{768}\) via the pretrained embedding lookup table from CodeT5p-220M, forming the embedding matrix \(\mathbf{E} = [\mathbf{e}_1, \mathbf{e}_2, \ldots, \mathbf{e}_m] \in \mathbb{R}^{m \times 768}\).}


In Step \circled{3}, the embedding matrix \(\mathbf{E}\) is fed into a T5 encoder student model \(\mathcal{S}_\theta\), consisting of 12 stacked encoder layers, each comprising layer normalization, multi-head self-attention, and a feed-forward sublayer \cite{raffel2020exploring}.
The final encoder representation is projected into a logit vector 
\(\mathbf{z} \in \mathbb{R}^{2}\), corresponding to the binary decision space \(\{\text{TP}, \text{FP}\}\). 

Let \(y \in \{0,1\}\) denote the pseudo-label provided by the LLM teacher \(\mathcal{T}\), where \(y=1\) indicates a false positive (FP) and \(y=0\) indicates a true positive (TP). We optimize the student model parameters \(\theta\) by minimizing the cross-entropy loss:
\[
\mathcal{L}_{\text{CE}}(\theta) = - \big[ y \cdot \log \sigma(\mathbf{z})_1 + (1-y) \cdot \log \sigma(\mathbf{z})_0 \big],
\]
where \(\mathbf{z} \in \mathbb{R}^2\) is the logit vector with \(z_0\) and \(z_1\) corresponding to TP and FP, respectively, and \(\sigma(\mathbf{z}) = [\sigma(\mathbf{z})_0, \sigma(\mathbf{z})_1]\) denotes the softmax function applied element-wise: \(\sigma(\mathbf{z})_k = \frac{e^{z_k}}{e^{z_0} + e^{z_1}}\) for \(k \in \{0,1\}\).
This objective distills the teacher’s labeling knowledge—its ability to discriminate false positives from \glitch—into the student model.
Through backpropagation and iterative updates of \(\theta\), the student model converges, inheriting the teacher’s ability to detect false positives in \glitch~warnings while using fewer parameters for improved efficiency.

\subsection{End-to-End Inference with \ourapp}
Given an IaC file \(x_\text{test}\) from the testing oracle, a security engineer first runs \glitch~to detect potential security smells using symbolic rules.
If \glitch~returns a warning \(w=(\ell, s) \in \mathcal{W}(x_\text{test})\), we extract the flagged line \(\ell\) and record its associated smell type \(s\). The flagged line alone is then tokenized, embedded, and passed through the trained student model \(\mathcal{S}_\theta\), which filters out false positives. Only warnings predicted as true positives by \ourapp~are reported back to security engineers.
\section{Experimental Design}
\label{sec:exp_design}
In this section, we present the motivation for our three research questions, our studied dataset, baseline approaches, and our experimental setup.

\subsection{Research Questions}
To evaluate our \ourapp~approach, we formulate the following three research questions.

\rqone
~Saavedra \ea~\cite{saavedra2022glitch} introduced \glitch, a state-of-the-art static analyzer for detecting security smells in IaC scripts. However, our preliminary analysis shows that \glitch~suffers from a high false alarm rate due to its reliance on symbolic pattern matching, a limitation common to static analysis tools \cite{charoenwet2024empirical}.
To address this issue, we propose \ourapp, which integrates symbolic rules with neural inference to reduce false positives. Therefore, we evaluate whether \ourapp~achieves higher accuracy than existing state-of-the-art approaches for IaC security smell detection.

\rqtwo
~One of the primary objectives of security smell detection in IaC scripts is to assist security engineers in locating problematic configurations in a cost-effective manner, i.e., uncovering the maximum number of true security smells with the least inspection effort. In practice, the effort required to review static analyzer warnings is a critical concern, as high false alarm rates can overwhelm engineers and hinder the adoption of approaches.
While accuracy is important, practical deployment also depends on whether an approach reduces the manual burden of triaging false positives. Therefore, we investigate the cost-effectiveness of our \ourapp~approach for locating security smells in IaC scripts, extending the evaluation beyond predictive accuracy to include inspection effort.

\rqthree
~Our \ourapp~approach incorporates several key design choices, including the selection of the LLM teacher model for generating pseudo labels during knowledge distillation, the student model type, and scale.
However, little is known about the impact of different design choices on the effectiveness of \ourapp~for IaC security smell detection.
In particular, the accuracy of pseudo-labels from the teacher model directly affects the supervision signal for distillation, while architectural choices and parameter scale influence the student model’s capacity and efficiency.
Thus, we formulate this RQ to investigate how alternative design choices influence the effectiveness of \ourapp, by varying each factor independently while keeping the others fixed.

\subsection{Studied Dataset}
To compare our \ourapp~approach with existing static analysis baselines for security smell detection, we require a test oracle that provides line-level labels indicating the presence and type of security smells in IaC scripts.
Additionally, to train \ourapp~to identify false positives produced by \glitch, we construct a separate dataset for knowledge distillation. Below, we detail each dataset.

\subsubsection{Test Oracles for IaC Security Smell Detection}
We use the same oracle test sets covering three IaC technologies—Puppet, Ansible, and Chef—provided by Saavedra \ea~\cite{saavedra2022glitch} to ensure a fair comparison with \glitch.
The Ansible oracle was collected from Rahman \ea's~81-script dataset~\cite{rahman2019seven}, while the Chef and Puppet oracles each comprise 80 scripts curated by Saavedra \ea~according to their selection criteria.
Each script may contain zero or more security smells. Seven independent raters annotated the scripts by identifying the smell type and corresponding line number without using any static analyzers.
In total, the oracle includes 65 labeled smell instances for Puppet, 44 for Ansible, and 105 for Chef, each represented as a \texttt{(line, type)} tuple. Additionally, the dataset contains 52 scripts labeled as smell-free for Puppet, 69 for Ansible, and 43 for Chef.
To ensure the reliability of the test oracle, we conducted an additional manual review of all 241 IaC scripts, including 214 security smells. We identified and removed one duplicate smell instance from the original labels, resulting in 213 security smells in total.

\subsubsection{Data for False Positive Detection}
To evaluate our student model on the false positive detection of \glitch's~warnings, we construct a test dataset by running \glitch\ on the test oracle scripts and labeling each detection as TP/FP by direct comparison to the oracle’s line-level ground truth.
Per-technology totals are Puppet 107, Ansible 48, and Chef 75 detections.
Next, we built training data using the replication package from Saavedra \ea~\cite{saavedra2022glitch}, which contains a large corpus of GitHub repositories for each IaC technology.
For each technology (Puppet, Ansible, Chef), we curated candidate pools of $\sim$2,000 scripts ("oracle-2000") by sampling from this corpus. To enrich positive examples and keep analysis tractable, we retained only files with (i) $\geq$20 \glitch-reported smell candidates and (ii) $\leq$200 lines. We applied MD5-based file-level deduplication to ensure no training candidate file was identical to any oracle test file.
The learning unit is a line of code. Each instance includes the target line, a contextual window of surrounding lines, and metadata (line number, smell type). An LLM teacher (Claude Sonnet 4.0) labeled candidate lines as TP/FP; these pseudo-labels then supervised a compact student model in our knowledge distillation process.

\smallsection{Data Deduplication Against Test Oracle}
We computed split sizes by working backward from our test dataset (Puppet 107, Ansible 48, Chef 75; total 230) using an 8:1 train-to-validation ratio, yielding targets of 1,840 train and 230 dev instances. To prevent leakage, we performed snippet-level deduplication (exact match on normalized code + smell type) with priority oracle $>$ validation $>$ train. After removing duplicates, we iteratively backfilled from candidate pools until all per-technology, per-smell targets were met, resulting in final train/validation counts of 856/107 for Puppet, 384/48 for Ansible, and 600/75 for Chef.

\subsection{Baselines}
We evaluate our approach against two sets of baselines for IaC security smell detection. The first set includes static analyzers: \textsc{SLIC}~\cite{rahman2019seven}, \textsc{SLAC}~\cite{rahman2021security}, and \glitch~\cite{saavedra2022glitch}. \textsc{SLIC} and \textsc{SLAC} are early rule-based tools tailored to specific IaC technologies, while \glitch~introduces a technology-agnostic intermediate representation for unified detection across Puppet, Ansible, and Chef. The second set comprises zero-shot LLMs—Claude 4.0, Grok 4, and GPT-5—prompted to identify security smells.
Each prompt comprises four components: system role, security smell description, IaC script input, and a structured output template. To populate the security smell description component, we adopt two prompting strategies. The first uses definition-based prompts, which provide descriptive explanations to help the LLM understand the semantics of each smell. The second employs rule-based prompts, encoding \glitch's symbolic rules and heuristics to guide the LLM in reasoning about detection. This results in six LLM-based trials (three models × two strategies).

\subsection{Experimental Setup}

\subsubsection{Targeted False Positive from \glitch}
\revised{Among the nine security smell types supported by \glitch, four—\textit{Hard-coded secret}, \textit{Suspicious comment}, \textit{Use of HTTP without TLS}, and \textit{Use of weak cryptographic algorithm}—exhibit comparatively lower detection precision. Our neural inference component is applied only to these four smell types to filter false positives, as \glitch~already achieves high precision on the remaining five types.}

\begin{figure*}[htbp!]
\includegraphics[width=\textwidth]{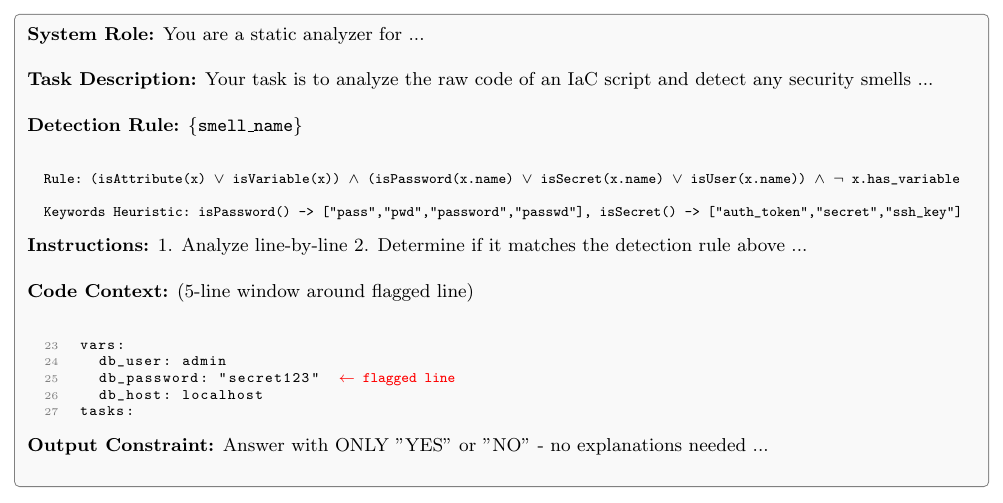}
\caption{Prompt template for pseudo-label generation.}
\label{fig:prompt-template}
\end{figure*}

\subsubsection{LLM Prompt Design for Generating Pseudo-Labels}
We use Claude Sonnet 4.0 as our teacher model to generate pseudo-labels for static analyzer false positive detection.
Figure~\ref{fig:prompt-template} presents our prompt template.
Our prompt template has three parts:
(i) a system role that frames the model as an IaC security analyst;
(ii) a task description that supplies the symbolic rule name and match rationale (i.e., predicates, and keywords heuristics) from \glitch, plus a 5-line code window around the target line;
and (iii) output constraints requiring a structured response. The exact template is available in our repository.
To ensure reproducibility, we enforce deterministic decoding (temperature=0, top\_p=1).


\subsubsection{Model Implementation and Training}
All models are implemented using HuggingFace Transformers (v4.45.2) and PyTorch (v2.4.0).
For the model architecture, we use ``Salesforce/codet5p-220m'' (220M parameters) with a maximum sequence length of 512 tokens.
We optimize the cross-entropy loss as described in Section \ref{subsec:ce_loss} for binary TP/FP classification.
The model was fine-tuned on NVIDIA A100 GPUs (40GB) on an HPC cluster.
During training, the learning rate is set to $4{\times}10^{-5}$ with cosine decay and 500 warmup steps. We use backpropagation with the AdamW optimizer (weight decay 0.01).
We monitor the validation set throughout training and select the checkpoint with the optimal F1-score.

\section{Experimental Results}
\label{sec:exp_results}
In this section, we present the experimental results addressing our three research questions.

\subsection*{\rqone}
\smallsection{Approach}
To address this RQ, we focus on detecting security-vulnerable code lines, commonly referred to as security smells, within IaC scripts across three widely used technologies: Puppet, Ansible, and Chef.
We compare our \ourapp~approach against two rule-based static analysis baselines (\glitch~and SLIC/SLAC), and three LLMs—Claude-4, Grok-4, and GPT-5—using two distinct prompting strategies: (i) definition-based prompts that provides descriptive definitions for security smells, and (ii) static analysis rule-based prompts that specify explicit detection patterns (i.e., symbolic rules, predicates, and keywords heuristics) defined in \glitch.
Each approach is evaluated on the oracle test set provided by Saavedra \ea~\cite{saavedra2022glitch}.
Each method scans the IaC scripts in the oracle test set one by one. Each script may contain zero or more security smells.
For each script, each method analyzes the entire script and outputs the lines identified as security smells, along with their corresponding smell types.
Following Saavedra \ea~\cite{saavedra2022glitch}, we use precision ($\frac{TP}{TP+FP}$) and recall ($\frac{TP}{TP+FN}$) as evaluation metrics.
To assess overall effectiveness, we also include the F1 score ($\frac{2\times\mathrm{Precision}\times\mathrm{Recall}}{\mathrm{Precision}+\mathrm{Recall}}$), which balances the trade-off between detecting more security smells and minimizing false positives.

\revised{Note that our results use micro-averaging and exclude scripts labeled as "No smell" from the oracle, whereas \glitch paper reports macro-averaged results including clean scripts.}

We also report the Macro-F1 score across the three technologies, defined as $\frac{1}{3} \left( F1_{\text{Puppet}} + F1_{\text{Ansible}} + F1_{\text{Chef}} \right)$, to provide a holistic view of performance.
Below, we define true positives (TP), false positives (FP), and false negatives (FN) used in computing our evaluation metrics:
\begin{itemize}
  \item \textbf{TP}: a predicted \texttt{(line, type)} exactly matches an oracle instance.
  \item \textbf{FP}: a predicted \texttt{(line, type)} does not exist in the oracle.
  \item \textbf{FN}: an oracle \texttt{(line, type)} has no matching prediction. 
\end{itemize}
\begin{table*}[t]
  \centering
  \caption{RQ1 — Detection performance on the oracle test set by technology (Puppet, Ansible, Chef), plus Macro-F1 (unweighted mean across the three).} 
  \label{tab:rq1}
  \scriptsize
  \setlength{\tabcolsep}{4pt}
  \resizebox{\linewidth}{!}{%
\begin{tabular}{l|ccc|ccc|ccc|c}
\hline
\multicolumn{1}{c|}{\multirow{2}{*}{\textbf{Method}}} & \multicolumn{3}{c|}{\textbf{Puppet}}                      & \multicolumn{3}{c|}{\textbf{Ansible}}                     & \multicolumn{3}{c|}{\textbf{Chef}}                        & \multirow{2}{*}{\textbf{Macro-F1}} \\ \cline{2-10}
\multicolumn{1}{c|}{}                        & \textbf{Prec}           & \textbf{Rec}            & \textbf{F1}             & \textbf{Prec}           & \textbf{Rec}            & \textbf{F1}             & \textbf{Prec}           & \textbf{Rec}            & \textbf{F1}             &                           \\ \hline
GLITCH                                       & 0.418          & \textbf{0.938} & 0.578          & 0.741          & \textbf{0.977} & 0.843          & 0.603          & \textbf{0.731} & 0.661          & 0.694                     \\
SLAC/SLIC                                    & 0.275          & 0.508          & 0.357          & 0.471          & 0.545          & 0.505          & 0.367          & 0.385          & 0.376          & 0.413                     \\ \hline
Claude-4 (rules)                             & 0.500          & 0.615          & 0.552          & 0.508          & 0.750          & 0.606          & 0.457          & 0.606          & 0.521          & 0.559                     \\
Claude-4 (def)                               & 0.623          & 0.738          & 0.676          & 0.610          & 0.818          & 0.699          & 0.490          & 0.490          & 0.490          & 0.622                     \\
Grok-4 (rules)                               & 0.718          & 0.785          & 0.750          & 0.833          & 0.909          & 0.870          & 0.667          & 0.635          & 0.650          & 0.757                     \\
Grok-4 (def)                                 & 0.667          & 0.738          & 0.701          & 0.857          & 0.682          & 0.759          & 0.625          & 0.577          & 0.600          & 0.687                     \\
GPT-5 (rules)                                & 0.542          & 0.692          & 0.608          & 0.678          & 0.909          & 0.777          & 0.554          & 0.596          & 0.574          & 0.653                     \\
GPT-5 (def)                                  & 0.593          & 0.538          & 0.565          & 0.741          & 0.455          & 0.563          & 0.655          & 0.346          & 0.453          & 0.527                     \\ \hline
\ourapp~(ours)  & \textbf{0.846} & 0.846          & \textbf{0.846} & \textbf{0.947} & 0.818          & \textbf{0.878} & \textbf{0.877} & 0.683          & \textbf{0.768} & \textbf{0.831}            \\ \hline
\end{tabular}
  }
  \vspace{0.25em}
  \begin{minipage}{0.98\textwidth}
    \footnotesize
    \emph{Abbreviations:} Claude-4 = Claude-Sonnet-4.0; Grok-4 = Grok-4-fast; GPT-5 = GPT-5-2025-08-07;
    “rules” uses our static-analysis-rules prompt; “def” uses the definition-based prompt. Best F1 per tech (and best Macro-F1) in \textbf{bold}.
  \end{minipage}
\end{table*}

\smallsection{Results}
Table~\ref{tab:rq1} presents per-technology results (Puppet, Ansible, and Chef) for our \ourapp~approach, two static analysis baselines, and six LLM-based methods, along with an aggregated Macro-F1 score across the three technologies.
\textbf{Our \ourapp~approach achieves the highest overall Macro-F1 of 83\%, outperforming static analysis methods (\glitch~and SLAC/SLIC) by 14\%–42\% and LLM-based methods by 7\%–30\%.}
Notably, our \ourapp~approach achieves the highest F1 across all three technologies and substantially improves upon \glitch, with improvement of 27\% on Puppet, 4\% on Ansible, and 11\% on Chef.
The observed performance gains are driven by substantial enhancements in precision, including a +43\% improvement on Puppet (42\%$\rightarrow$85\%), a +21\% improvement on Ansible (74\%$\rightarrow$95\%), and a +28\% improvement on Chef (60\%$\rightarrow$88\%).
Importantly, these gains come at only modest recall reductions (\textminus9\%, \textminus16\%, and \textminus5\%, respectively).
For static analysis methods, \glitch~maintains the highest recall, but its lower precision limits overall performance.
SLAC/SLIC consistently underperforms relative to other baselines.
Among LLM-based methods, the two prompting styles yield comparable results; notably, grok-4-fast with static-analysis-rules prompting performs best (Macro-F1 of 76\%), surpassing \glitch~on Puppet and Ansible.
\textbf{These experimental results confirm that our \ourapp~approach is the most accurate for detecting security smells in IaC scripts, driven by substantial precision enhancements that address the limitation of \glitch~and yield the highest F1 scores.}

\subsection*{\rqtwo}
\smallsection{Approach}
To address this RQ, we evaluate the cost-effectiveness of our \ourapp~approach for detecting security smells in IaC scripts.  
In a practical adoption scenario, the most cost-effective detection methods are those that enable security engineers to identify the greatest number of actual smells with the least inspection effort.  
For this evaluation, we assume that the 241 scripts in the testing oracle (i.e., 11,814 lines of code) represent the set of IaC scripts that security engineers would need to review.
To evaluate the cost-effectiveness of our \ourapp~approach, we first obtain its predictions and then rank them using the raw sigmoid output scores produced by our false positive pruner.
Lines with higher scores, which indicate a greater likelihood of being security smells, are placed at the top of the ranking.
In practice, this ranking helps security engineers prioritize inspection, focusing on the most suspicious code first and reducing the effort needed to find actual security smells in IaC scripts.
We evaluate the cost-effectiveness using two effort-aware metrics:
\begin{enumerate}
\item[(1)] \textbf{Effort@60\%Recall} measures the amount of effort (measured as LOC) that security analysts have to spend to find out the actual 60\% security smells. It is computed as the total LOC used to locate 60\% of the security smells divided by the total LOC in the testing oracle. A lower value of Effort@60\%Recall indicates that the security analysts may spend a smaller amount of effort to find the 60\% actual security smells. Note that methods that cannot achieve this recall are marked \emph{unreached}.
\item[(2)] \textbf{F1@1\%LOC} measures the balance between precision and recall when evaluating how effectively security smells are identified within a fixed amount of effort (i.e., the top 1\% of LOC in the testing oracle).
It is computed as the harmonic mean of precision and recall, where precision is the proportion of correctly predicted security smells among all smells in the top 1\%, and recall is the proportion of actual security smells retrieved within the top 1\%.
A higher value of F1@1\%LOC indicates that an approach not only ranks many actual security smells at the top but also minimizes wasted effort on false positives, thus providing a more cost-effective measure of security smell detection.
\end{enumerate}
We use the same testing oracle across the three IaC technologies—Puppet, Ansible, and Chef—and compare against the same set of baseline methods as in RQ1.
Because the LLM APIs we use (GPT-5, Grok-4, and Claude-4) do not return confidence scores, we prompt each LLM baseline again to assign a confidence score in the range $[0, 1]$ for its previously detected security smells.
For the rule‑based static analyzers (\glitch~and SLAC/SLIC), which do not produce confidence scores, we assign a constant value of 1.0 to all alerts. We then use the confidence scores to rank the detections.



\begin{table}[htbp!]
  \centering
  \caption{(RQ2 Result) The Effort@60\%Recall and F1@1\%LOC of our \ourapp~approach and eight baseline approaches. ($\searrow$) Lower Effort@60\%Recall = Better, ($\nearrow$) Higher F1@1\%LOC = Better.}
  \label{tab:rq2-compact}
  \footnotesize
  \setlength{\tabcolsep}{3.5pt}
  \renewcommand{\arraystretch}{1.05}
  \resizebox{\columnwidth}{!}{%
\begin{tabular}{l|cc|cc|cc}
\hline
\multicolumn{1}{c|}{\multirow{2}{*}{\textbf{Method}}} & \multicolumn{2}{c|}{\textbf{Puppet}}    & \multicolumn{2}{c|}{\textbf{Ansible}}   & \multicolumn{2}{c}{\textbf{Chef}}       \\ \cline{2-7} 
\multicolumn{1}{c|}{}                        & \textbf{Eff@60\%}      & \textbf{F1@1\%}         & \textbf{Eff@60\%}      & \textbf{F1@1\%}         & \textbf{Eff@60\%}      & \textbf{F1@1\%}         \\ \hline
GLITCH                                       & 2.40          & 0.343          & 0.96          & 0.707          & 2.57          & 0.359          \\
SLAC/SLIC                                    & ---           & 0.248          & ---           & 0.439          & ---           & 0.262          \\ \hline
Claude-4 (rules)                             & 1.29          & 0.610          & 0.98          & 0.659          & ---           & 0.469          \\
Claude-4 (def)                               & 1.34          & 0.533          & 0.82          & 0.829          & ---           & 0.428          \\
Grok-4 (rules)                               & 1.19          & 0.610          & 0.77          & \textbf{0.854} & 2.15          & 0.400          \\
Grok-4 (def)                                 & 1.24          & 0.571          & 0.80          & 0.683          & ---           & 0.414          \\
GPT-5 (rules)                                & 1.72          & 0.400          & 0.93          & 0.732          & ---           & 0.331          \\
GPT-5 (def)                                  & ---           & 0.476          & ---           & 0.439          & ---           & 0.359          \\ \hline
\ourapp~(ours)  & \textbf{1.14} & \textbf{0.648} & \textbf{0.74} & \textbf{0.854} & \textbf{1.66} & \textbf{0.552} \\ \hline
\end{tabular}
  }
  \vspace{0.25em}
  \begin{minipage}{\columnwidth}
    \footnotesize
    \emph{Notes:} “---” denotes \emph{unreached} (method’s $max\_recall<60\%$). Best per column in \textbf{bold} (ties are all bolded).
  \end{minipage}
\end{table}

\smallsection{Results}
Table~\ref{tab:rq2-compact} presents the experimental results of our \ourapp~approach and the eight baseline approaches measured using Effort@60\%Recall and F1@1\%LOC.

\textbf{Our \ourapp~consistently outperforms all baselines, achieving best Effort@60\%Recall of 1.14\%, 0.74\%, and 1.66\% on Puppet, Ansible, and Chef.}
On Puppet, our \ourapp~requires 1.14\% LOC, representing 4–53\% less effort (two baselines failed to reach 60\% recall).
On Ansible, our \ourapp~requires 0.74\% LOC, corresponding to 4–25\% less effort compared with other baselines (two baselines failed to reach 60\% recall). 
On Chef, our \ourapp~requires 1.66\% LOC, yielding 23–35\% less effort (six baselines failed to reach 60\% recall). 
The relative improvement is calculated as $(\text{baseline} - \text{ours}) / \text{baseline}$, since lower LOC is better.
In practical terms, these Effort@60\%Recall values indicate that security engineers need to inspect only 1.14\% of the Puppet code base ($\approx$45 LOC out of 3,964), 0.74\% of the Ansible code base ($\approx$28 LOC out of 3,760), and 1.66\% of the Chef code base ($\approx$68 LOC out of 4,090) to detect 60\% of the security smells in each code base.
This finding implies that \ourapp~can help security engineers detect the same number of security smells with less inspection effort.

\textbf{In terms of F1@1\%LOC, our \ourapp~achieves the best measure of 65\%, 85\%, and 55\% on Puppet, Ansible, and Chef.}
On Puppet, \ourapp~achieves an F1@1\%LOC of 65\%, which is 6-161\% better than other baselines and comparable to the best baseline but with better Effort@60\%Recall. 
On Chef, \ourapp~achieves an F1@1\%LOC of 55\%, corresponding to an 18–111\% improvement over other baselines. 
On Ansible, \ourapp~achieves 85\%, yielding a 3-95\% improvement over other baselines.

The relative improvement is calculated as
$(\text{ours} \allowbreak - \allowbreak \text{baseline}) \allowbreak / \allowbreak \text{baseline})$.
Concretely, when security engineers review only the top 1\% of lines flagged by \ourapp ($\approx$40 in Puppet, 38 LOC in Ansible, and 41 in Chef), the resulting detection quality—measured by F1@1\%LOC—is 65\% on Puppet, 85\% on Ansible, and 55\% on Chef, indicating strong precision–recall balance even when engineers can only review a very small fraction of the code.
This finding suggests that our \ourapp~may help security engineers achieve higher detection quality by identifying more true security smells with fewer false alarms than baseline approaches, given the same inspection effort.

\textbf{In summary, these experimental results confirm that \ourapp~is highly cost-effective for locating security smells in IaC scripts. It consistently reduces the inspection effort required to detect the same proportion of security smells and delivers higher detection quality under the same inspection effort.}


\subsection*{\rqthree}
\smallsection{Approach}
To address this RQ, we investigate three key design choices in our \ourapp~approach: (1) the LLM Teacher Model used during knowledge distillation, (2) the underlying model type, and (3) the model scale. For each experiment, we evaluate performance using the same oracle testing set as in RQ1 and RQ2.

To assess the impact of LLM Teacher Model choice, we compare Claude-4 (Anthropic) with two alternatives, GPT-5 (OpenAI) and Grok-4 (XAI), while keeping the model type and scale fixed to CodeT5p-220M.
We conduct two complementary evaluations using the same oracle test set.
First, we measure the labeling accuracy of each LLM in identifying false positives among static analysis warnings produced by \glitch, using a consistent prompt format that includes IaC code, warning context, and rule metadata used in \glitch. This experiment reflects the quality of supervision used during distillation.
Second, we retrain \ourapp~using training sets labeled by each LLM and evaluate downstream performance on the IaC security smell detection task. For both evaluations, we report precision, recall, and F1 score to quantify the effect of LLM Teacher Model choice on labeling accuracy and detection effectiveness.

To evaluate the impact of model type and scale, we conduct two controlled experiments using a fixed labeling source (Claude-4).
First, we compare our CodeT5p backbone with three alternative pretrained models: CodeT5, CodeBERT, and UniXcoder
We use comparable parameter scales across variants to isolate architectural effects.
Second, we assess the influence of model size by comparing CodeT5p-220M and CodeT5p-770M, holding both the LLM Teacher Model and model type constant.
For both experiments, we report macro-F1 scores across three IaC technologies (Puppet, Ansible, Chef), capturing overall effectiveness in detecting security smells while minimizing false alarms.

\begin{table}[htbp!]
\centering
\caption{(RQ3 - False Positive Detection) Precision, recall, and F1 scores for false positive detection on \glitch~warnings using three LLM pseudo-labelers.}
\label{tab:rq3_1_postfilter}
\footnotesize
\setlength{\tabcolsep}{3.5pt}
\renewcommand{\arraystretch}{1.05}
\resizebox{0.7\columnwidth}{!}{%
\begin{tabular}{l|ccc}
\hline
Teacher Model & Precision & Recall & F1 \\
\hline
GPT-5 & \textbf{0.952} & 0.692 & 0.802 \\
Grok-4 & 0.894 & 0.531 & 0.667 \\
Claude-4 (ours) & 0.924 & \textbf{0.853} & \textbf{0.887} \\
\hline
\end{tabular}
}
\vspace{0.25em}
\begin{minipage}{\columnwidth}
\footnotesize
\emph{Notes:} Best value per column in \textbf{bold}.
\end{minipage}
\end{table}

\begin{table}[htbp!]
  \centering
  \caption{(RQ3 - Downstream Security Smell Detection) Precision, recall, and F1 scores of our \ourapp~for security smell detection using three different LLM pseudo-labelers.}
  \label{tab:rq3_1_downstream_refactor}
  \footnotesize
  \setlength{\tabcolsep}{6pt}
  \renewcommand{\arraystretch}{1.05}
  \resizebox{\columnwidth}{!}{%
  \begin{tabular}{l|ccc}
    \hline
    Method & Precision & Recall & F1 \\
    \hline
    GPT-5 + CodeT5p-220M      & $0.683$ & $0.759$ & $0.717$ \\
    Grok-4 + CodeT5p-220M     & $0.519$ & $ \textbf{0.914} $
 & $0.662$ \\
    Claude-4 + CodeT5p-220M~(ours) & $ \textbf{0.729} $ & $0.730$ & $ \textbf{0.727} $ \\
    \hline
  \end{tabular}
  }
  \vspace{0.25em}
  \begin{minipage}{\columnwidth}
    \footnotesize
    \emph{Notes:} Best value per column in \textbf{bold}.
  \end{minipage}
\end{table}

\begin{table}[htbp!]
  \centering
  \caption{(RQ3 - Model Type and Scale) The performance of security smell detection when using different model types and scales in our \ourapp~approach.}
  \label{tab:rq3_2_scaling_lite}
  \footnotesize
  \setlength{\tabcolsep}{6pt}
  \renewcommand{\arraystretch}{1.05}
  \resizebox{\columnwidth}{!}{%
  \begin{tabular}{l|c|c}
    \hline
    Student Model & Params (M) & Macro F1 (argmax) \\
    \hline
    CodeBERT\textendash base    & 110 & 0.766 \\
    UniXcoder\textendash base   & 110 & 0.746 \\
    CodeT5\textendash base      & 220 & 0.769 \\
    CodeT5p\textendash 220M (ours)     & 220 & \textbf{0.794} \\
    \hline
    CodeT5p\textendash 770M     & 770 & 0.785 \\
    CodeT5p\textendash 220M (ours)     & 220 & \textbf{0.794} \\
    \hline
  \end{tabular}
  }
  \vspace{0.25em}
  \begin{minipage}{\columnwidth}
    \footnotesize
    \emph{Notes:} Best Macro F1 in \textbf{bold}.
  \end{minipage}
\end{table}

\smallsection{Results}
Table~\ref{tab:rq3_1_postfilter} presents the false positive detection results for three LLM Teacher Models.
GPT-5 achieves higher precision (95\%) than Claude-4 (92\%), indicating stronger retention of true positive warnings of \glitch.
Claude-4 has the highest recall (85\%), which reflects a better ability to flag false positive warnings of \glitch, compared to GPT-5 (69\%) and Grok-4 (53\%).
Overall, Claude-4 offers the best balance between preserving true positives and identifying false positives with an F1 score of 89\%, making it the most effective labeler for distillation.

Table~\ref{tab:rq3_1_downstream_refactor} presents the security smell detection results when retraining CodeT5p-220M in our \ourapp~using training sets labeled by three different LLM Teacher Models.
Models trained with GPT-5 and Grok-4 labels achieve higher recall (76\% and 91\%) but lower precision (68\% and 52\%), indicating stronger retention of true positives of \glitch~warnings but limited false positive filtering.
In contrast, Claude-4 labels yield balanced precision and recall (73\% each) and the highest F1 score (73\%), demonstrating a better effectiveness in filtering out the false positive warnings from \glitch.
This outcome is consistent with the expected behavior of student models inheriting the labeling characteristics of their teacher LLMs during distillation.

Table~\ref{tab:rq3_2_scaling_lite} presents the security smell detection results across different model types and scales. CodeT5p achieves the highest macro-F1 score (79\%), confirming its effectiveness as our chosen backbone. When varying model scale, both CodeT5p-220M and CodeT5p-770M yield comparable macro-F1 scores (79\%), but our model is 71\% smaller in terms of parameter count, confirming our design choice for computational efficiency without sacrificing performance.
\revised{The comparable performance between 220M and 770M variants can be attributed to our relatively small training set (2,070 samples), where both models converge well and stochastic training variations dominate.}

\vspace{-2mm}
\section{Discussion}
\label{sec:discussion}
In this section, we analyze \ourapp’s results across the nine security smell types, as well as its performance on scripts without any smells (No Smells) in the test oracle. Table~\ref{tab:oracle-ours-per-tech} reports the precision, recall, and F1 scores achieved across Puppet, Ansible, and Chef.

\ourapp~performs strongly on several smells, with precision and recall both above 80\% for \textit{Invalid IP address binding} (achieving a perfect prediction), \textit{Empty passwords}, \textit{Suspicious comments}, and \textit{Missing default case statements}. For others, including \textit{Admin by default}, \textit{Use of HTTP without TLS}, and \textit{Hard-coded secrets}, performance remains promising, though in the case of \textit{Hard-coded secrets}, precision improves to 79\% while recall drops to 51\%, reflecting a trade-off when applying neural inference on top of traditional static analysis to reduce false positives.
Additionally, our \ourapp~achieves a perfect precision for \textit{No smell}, indicating that it reliably avoids false positives when classifying clean scripts, highlighting the effectiveness of the neural inference component in \ourapp.

Performance is lower for \textit{Use of weak cryptographic algorithms} and \textit{No integrity check}.
These smells occur infrequently in our data, which limits the number of training samples and results in unstable performance of \ourapp. Since our training set reflects the natural distribution of security smells in real-world data, the model had fewer opportunities to learn the patterns and reduce false positives from \glitch~in the case of \textit{Use of weak cryptographic algorithms}. For \textit{No integrity check}, the static analyzer provides limited coverage of this smell type, leading to suboptimal detection.

These findings suggest that while \ourapp~is effective in detecting several prevalent security smells, its performance is less consistent for infrequent or weakly supported types.
For researchers, the findings point to several opportunities for improvement: enhancing the neural inference component of \ourapp, expanding or rebalancing training datasets through data augmentation to better represent rare smell types, and refining symbolic rule matching—particularly for underrepresented cases such as \textit{No integrity check}.
For security practitioners, \ourapp~can serve as a useful tool for identifying common misconfigurations, but should be supplemented with manual inspection or complementary tools when addressing less frequent or complex smells.

\revised{Finally, regarding cost-effectiveness, our distilled student model (CodeT5p-220M) preserves 98.2\% of the teacher model’s F1 score while being approximately 500× smaller, enabling local deployment without API costs and only marginal performance loss.}

\begin{table}[htbp!]
  \centering
  \caption{(Discussion) The performance of our \ourapp~approach across different security smells.}
  \label{tab:oracle-ours-per-tech}
  \resizebox{\linewidth}{!}{%
  \begin{tabular}{l|c|ccc}
    \hline
    \textbf{Smell Name} & \textbf{Occurr.} & \textbf{Precision} & \textbf{Recall} & \textbf{F1} \\
    \hline
    Invalid IP address binding       & 14  & 1.00 & 1.00 & 1.00 \\
    Empty password                   & 10  & 1.00 & 0.92 & 0.95 \\
    Suspicious comment               & 19  & 0.97 & 0.83 & 0.87 \\
    Missing default case statement   & 30  & 0.92 & 0.97 & 0.94 \\ \hline
    Admin by default                 & 57  & 0.92 & 0.78 & 0.82 \\
    Use of HTTP without TLS          & 38  & 0.76 & 0.97 & 0.84 \\
    Hard-coded secret                & 32  & 0.79 & 0.51 & 0.62 \\ \hline
    Use of weak crypto alg.          & 5   & 0.53 & 1.00 & 0.70 \\
    No integrity check               & 8   & 0.33 & 0.33 & 0.33 \\
    \hline
    \textbf{Average}                 &  213  & 0.80 & 0.81 & 0.79 \\
    \textbf{Weighted Average}                 &  213  & 0.85 & 0.81 & 0.81 \\
    \hline
    No smell                         & 164 & 1.00 & 0.95 & 0.98 \\
    \hline
  \end{tabular}}
\end{table}

\vspace{-1mm}
\section{Threats to Validity}
\label{sec:threats}
\textbf{Threats to construct validity} relate to the accuracy of the human-labeled test oracle used for evaluation.
Although the oracle provided by Saavedra \ea~\cite{saavedra2022glitch} was carefully constructed—using closed coding by multiple raters with IaC and security expertise, independent annotations to reduce bias, and agreement checks across seven raters—manual labeling of security smells inherently involves subjectivity and potential human error.
To mitigate this threat, we manually reviewed the test oracle, including 241 IaC scripts with 214 labeled security smells.

\textbf{Threats to internal validity} relate to the hyperparameter settings used in the knowledge distillation of our \ourapp~framework. Knowledge distillation involves several tunable parameters, and the stochastic nature of backpropagation can introduce variability. We followed the default training configurations recommended in the original language model papers for both our main approach and the RQ3 variants. Since our goal is not hyperparameter optimization but building an efficient false-positive detector to complement traditional static analyzers, we consider this choice appropriate.
To mitigate this threat, we provide complete training scripts and configurations to ensure full reproducibility.

\textbf{Threats to external validity} relate to the generalizability of our \ourapp~approach across different IaC technologies, security smell types, and static analyzers. Our experiments, based on nine security smells from the human-labeled oracle of Saavedra \ea~\cite{saavedra2022glitch}, ensure a fair comparison with \glitch, the state-of-the-art analyzer our approach builds upon. However, the results may not fully generalize to other analyzers, additional smell types, or IaC technologies beyond Puppet, Ansible, and Chef.
Nonetheless, the \ourapp~framework is, in principle, applicable to broader contexts. To mitigate this threat, we release our implementation, dataset, and results as open source, enabling replication and extension. Future work should examine more IaC technologies, smell types, and analyzers to further validate and generalize our findings.
\revised{Finally, while \ourapp~substantially improves precision by filtering \glitch~alerts, this filtering may also remove some true positives, consistent with the modest recall reductions observed in RQ1. Moreover, \ourapp~cannot recover false negatives—security smells that \glitch~fails to detect in the first place. Addressing such false negatives would require enhancing \glitch’s symbolic rules or incorporating complementary detection techniques, a limitation that future work should explore.}


\section{Conclusion}
In this paper, we propose \ourapp, an intelligent static analyzer that integrates symbolic rule matching with neural inference to detect security smells in IaC scripts. We conduct an empirical evaluation on a human‑labeled dataset comprising 241 IaC scripts (11,814 lines of code) across three widely used technologies: Puppet, Ansible, and Chef. The results show that \ourapp~achieves the best detection performance, with F1 scores of 85\%, 88\%, and 77\% on Puppet, Ansible, and Chef, respectively. In terms of cost‑effectiveness, \ourapp~outperforms baselines, achieving the best Effort@60\%Recall (1.14\%, 0.74\%, and 1.66\%) and F1@1\%LOC (65\%, 85\%, and 55\%). An ablation study further validates the rationale behind our design choices.
Overall, the findings demonstrate that \ourapp~is both more accurate and more cost‑effective than existing static analyzers for IaC security smell detection, indicating its potential to assist security engineers in identifying insecure patterns within IaC at scale.


\bibliographystyle{ACM-Reference-Format}
\bibliography{reference}




\end{document}